\documentclass[%
 reprint,
superscriptaddress,
 amsmath,amssymb,
 aps,
prd
]{revtex4-2}

\usepackage{url}
\usepackage{xspace}

\usepackage{graphicx}
\usepackage{dcolumn}
\usepackage{bm}
\usepackage{hyperref}
\usepackage{gensymb}
\usepackage{booktabs}
\usepackage{xcolor, soul}
\sethlcolor{yellow}
\usepackage{cancel}
\begin{document}


\title{Measurement of the scintillation resolution in liquid xenon and its impact for future segmented calorimeters}


\author{C.~Romo-Luque}
\altaffiliation{Now at Los Alamos National Laboratory, US.}
\affiliation{Instituto de F\'isica Corpuscular (IFIC), CSIC \& Universitat de Val\`encia, Calle Catedr\'atico Jos\'e Beltr\'an, 2, Paterna, E-46980, Spain}
\author{N. Salor-Igui\~niz}
\email{Corresponding author: nerea.salor@dipc.org}
\affiliation{Donostia International Physics Center, BERC Basque Excellence Research Centre, Manuel de Lardizabal 4, Donostia-San Sebasti\'an, E-20018, Spain}
\affiliation{Instituto de F\'isica Corpuscular (IFIC), CSIC \& Universitat de Val\`encia, Calle Catedr\'atico Jos\'e Beltr\'an, 2, Paterna, E-46980, Spain}
\author{J.M.~Benlloch-Rodr\'{i}guez}
\affiliation{Donostia International Physics Center, BERC Basque Excellence Research Centre, Manuel de Lardizabal 4, Donostia-San Sebasti\'an, E-20018, Spain}
\author{R.~Esteve}
\affiliation{Instituto de Instrumentaci\'on para Imagen Molecular (I3M), Centro Mixto CSIC - Universitat Polit\`ecnica de Val\`encia, Camino de Vera s/n, Valencia, E-46022, Spain}
\author{R.J.~Aliaga}
\affiliation{ Instituto Universitario de Matem\'atica Pura y Aplicada (IUMPA), Universitat Polit\`ecnica de Val\`encia, Camino de Vera s/n, Valencia, E-46022, Spain}
\author{V.~\'Alvarez}
\affiliation{Instituto de Instrumentaci\'on para Imagen Molecular (I3M), Centro Mixto CSIC - Universitat Polit\`ecnica de Val\`encia, Camino de Vera s/n, Valencia, E-46022, Spain}
\author{F.~Ballester}
\affiliation{Instituto de Instrumentaci\'on para Imagen Molecular (I3M), Centro Mixto CSIC - Universitat Polit\`ecnica de Val\`encia, Camino de Vera s/n, Valencia, E-46022, Spain}
\author{R.~Gadea}
\affiliation{Instituto de Instrumentaci\'on para Imagen Molecular (I3M), Centro Mixto CSIC - Universitat Polit\`ecnica de Val\`encia, Camino de Vera s/n, Valencia, E-46022, Spain}
\author{J. Generowicz}
\affiliation{FULL BODY INSIGHT, S.L. Plaza Juan de Ribera 7-A, 46520, Puerto de Sagunto, Spain}
\author{A. Laing}
\affiliation{FULL BODY INSIGHT, S.L. Plaza Juan de Ribera 7-A, 46520, Puerto de Sagunto, Spain}
\author{A. Mart\'inez}
\affiliation{Instituto de F\'isica Corpuscular (IFIC), CSIC \& Universitat de Val\`encia, Calle Catedr\'atico Jos\'e Beltr\'an, 2, Paterna, E-46980, Spain}
\author{F.~Monrabal}
\affiliation{Donostia International Physics Center, BERC Basque Excellence Research Centre, Manuel de Lardizabal 4, Donostia-San Sebasti\'an, E-20018, Spain}
\affiliation{Ikerbasque (Basque Foundation for Science), Bilbao, E-48009, Spain}
\author{M.~Querol}
\affiliation{Instituto de F\'isica Corpuscular (IFIC), CSIC \& Universitat de Val\`encia, Calle Catedr\'atico Jos\'e Beltr\'an, 2, Paterna, E-46980, Spain}
\author{M. Rappaport}
\affiliation{Weizmann Institute of Science, Herzl St 234, Rehovot, Israel}
\author{J.~Rodr\'iguez}
\affiliation{Instituto de Instrumentaci\'on para Imagen Molecular (I3M), Centro Mixto CSIC - Universitat Polit\`ecnica de Val\`encia, Camino de Vera s/n, Valencia, E-46022, Spain}
\author{J.~Rodr\'iguez-Ponce}
\affiliation{Instituto de F\'isica Corpuscular (IFIC), CSIC \& Universitat de Val\`encia, Calle Catedr\'atico Jos\'e Beltr\'an, 2, Paterna, E-46980, Spain}
\author{S.~Teruel-Pardo}
\affiliation{Instituto de F\'isica Corpuscular (IFIC), CSIC \& Universitat de Val\`encia, Calle Catedr\'atico Jos\'e Beltr\'an, 2, Paterna, E-46980, Spain}
\author{J.F.~Toledo}
\affiliation{Instituto de Instrumentaci\'on para Imagen Molecular (I3M), Centro Mixto CSIC - Universitat Polit\`ecnica de Val\`encia, Camino de Vera s/n, Valencia, E-46022, Spain}
\author{R.~Torres-Curado}
\affiliation{Instituto de Instrumentaci\'on para Imagen Molecular (I3M), Centro Mixto CSIC - Universitat Polit\`ecnica de Val\`encia, Camino de Vera s/n, Valencia, E-46022, Spain}
\author{P.~Ferrario}
\altaffiliation{Also at Instituto de Física Corpuscular (Spain)}
 \altaffiliation{On leave.}
\affiliation{Donostia International Physics Center, BERC Basque Excellence Research Centre, Manuel de Lardizabal 4, Donostia-San Sebasti\'an, E-20018, Spain}
\affiliation{Ikerbasque (Basque Foundation for Science), Bilbao, E-48009, Spain}
\author{V.~Herrero}
\affiliation{Instituto de Instrumentaci\'on para Imagen Molecular (I3M), Centro Mixto CSIC - Universitat Polit\`ecnica de Val\`encia, Camino de Vera s/n, Valencia, E-46022, Spain}
\author{J.J.~G\'omez-Cadenas}
\affiliation{Donostia International Physics Center, BERC Basque Excellence Research Centre, Manuel de Lardizabal 4, Donostia-San Sebasti\'an, E-20018, Spain}
\affiliation{Ikerbasque (Basque Foundation for Science), Bilbao, E-48009, Spain}


\date{\today}

\begin{abstract}
We report on a new measurement of the energy resolution that can be attained in liquid xenon when recording only the scintillation light. Our setup is optimized to maximize light collection, and uses state-of-the-art, high-PDE, VUV-sensitive silicon photomultipliers. We find a value of $3.7 \pm 0.4$ $\%$  at 511 keV, once saturation effects are corrected for, a result close to the Poissonian resolution that we expect in our setup ($2.8 \pm 0.4\%$ at 511 keV). 

Our results in the intrinsic resolution ($2.3 \pm 0.8$ $\%$ ) are compatible, within errors, at 511 keV, with those found by theoretical estimations which have been standing for the last twenty years,  1.8 $\%$ . Our work opens new possibilities for apparatus based on liquid xenon and using scintillation only. In particular it suggests that modular scintillation detectors using liquid xenon can be very competitive as building blocks in segmented calorimeters, with applications to Positron Emission Tomography technology. 
\end{abstract}

\maketitle

\section{INTRODUCTION}

A large number of experiments in nuclear and particle physics require the measurement of energy, position and timing with high resolution. Good energy resolution can be attained with the use of high light-yield materials ($y$), while time resolution requires fast decay time of the scintillator ($\tau$). In addition, calorimetry requires high density and good optical quality, e.g., a material transparent to its own emission light. 

A number of crystals made of inorganic scintillators offer, to variable degrees, all of the above. Among these, the most extensively used is NaI(Tl), a very bright ($y = 38\,000$~photons/MeV) and fast scintillator ($\tau =250$~ns \cite{naI}). Other popular inorganic scintillators are CsI(Tl) ($y = 17\,100$~photons/MeV, $\tau =900$~ns \cite{csI}), LYSO(Ce) ($y = 25\,000$~photons/MeV, $\tau =40$~ns \cite{lysoCe} ) and BGO ($y\sim 9\,000$~photons/MeV, $\tau =300$~ns  \cite{bgo}).
The density of the above scintillators varies between  3.7 g/cm$^3$ (for NaI) and 7.2 g/cm$^3$ (for LYSO). Good spatial resolution is also often a requirement of nuclear and particle physics experiments. This can be achieved, for example, by segmenting the scintillators (along the transverse plane relative to the incident particles).

All of the above crystals have either been used in, or actively pursued for, high energy and nuclear physics experiments. NaI was used in Crystal Ball, CsI in CLEO and BGO in L3. LYSO is one of the best candidates for future collider applications, due to its excellent properties (high light yield, fast decay time, high density) and radiation hardness \cite{MaterialsFutureColliders}.
LYSO is also the preferred scintillation material in modern Positron Emission Tomography (PET)  scanners \cite{review_fb, Westerwoudt_2014, Conti_2019}. PET scanners benefit from a good energy resolution, because energy resolution allows one to reject a larger Compton scatter fraction, i.e., coincidences where one or both photons interact in the body of the patient and lose part of their energy. Those coincidences add noise to the image, since the direction of the gamma changes after the scattering.
 
\begin{figure}[htbp]
\centering 
\includegraphics[width=0.5\textwidth]{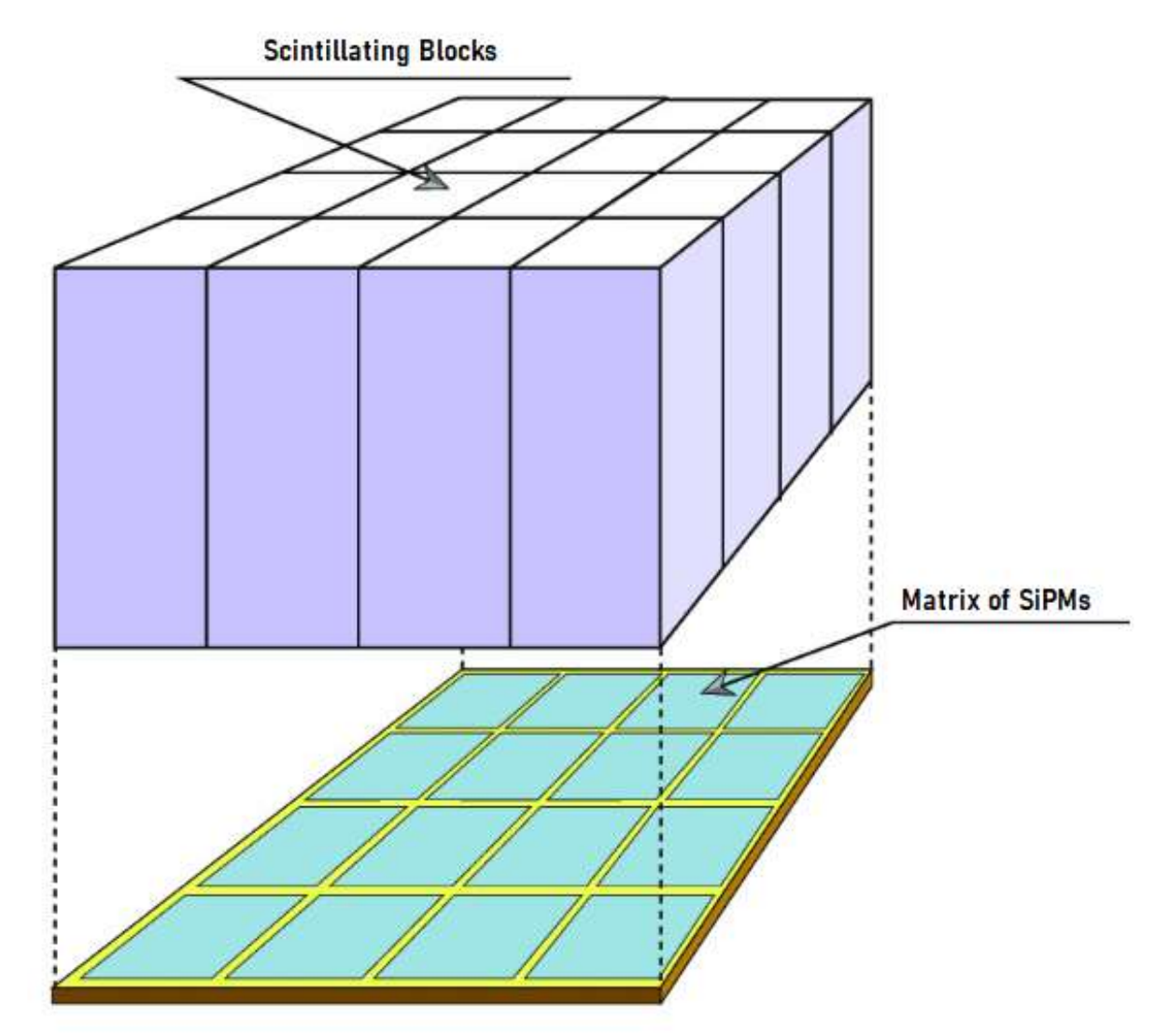}
\caption{\label{fig:ssb} A schematic illustrating the concept of SSB.}
\end{figure}

The common element in many nuclear and particle calorimeters and in modern PET scanners is a {\em Segmented Scintillating Block}, SSB (Fig. \ref{fig:ssb}). An SSB is an array of contiguous cells filled with scintillating material. The light propagates inside each cell by reflection, until it reaches the readout sensor, normally a SiPM in modern devices, although PMTs and APDs were used in the past. The thickness of the cells depends on the application and varies between one and a few attenuation lengths. The SSB can be characterized by three parameters: energy resolution, which depends on the scintillator's light yield, as well as the optical collection efficiency of the cell, the photodetection efficiency (PDE) of the SiPM, electronic noise, etc; time resolution, which is dominated by the scintillator's decay time; and (transverse) spatial resolution, which can be tailored by choosing a suitable transverse size for the cell. 
As an example, in the 2 m long EXPLORER scanner \cite{Cherry_2018, Badawi_2019}, the largest reported PET, the SSB material is LYSO, with cell area $2.76 \times 2.76$ mm$^2$ and thickness of 1.6 attenuation lengths, and the cells are read out by SiPMs. The number of individual cells exceeds $500\, 000$. The energy resolution at 511 keV is 12\% FWHM, the spatial resolution $\sim$3 mm and the time-of-flight (TOF) resolution is 500 ps. 
  
 Liquid xenon (LXe) is also an excellent calorimeter \cite{Aprile:2010}. 
 Charged particles interact with LXe via both ionization and excitation. The excitation of xenon atoms results in strongly bound excited molecules (excimers) which, on de-excitation, emit VUV photons with a wavelength of  $\sim$175 nm \cite{FUJII2015293}.
 In addition, the xenon atoms are ionized. An external electric field can be used to carry away the ionization electrons, in which case two signals are available, one from scintillation, and one from ionization. In the absence of such electric field, most ionization electrons eventually recombine with positive xenon ions, and both channels (primary scintillation and recombination) contribute to the resulting scintillation yield of $\sim 58\, 700$ photons per MeV \cite{Chepel:2012sj}, which is a factor 1.5 larger than NaI and 1.7 larger than LYSO. Several measurements have found yield fractions of primary scintillation and recombination of roughly 0.26 and 0.74 for electrons of around 1 MeV of energy \cite{PhysRevB.17.2762}. The primary scintillation is characterised by two decay constants ($\tau_1 = 3$ ns, $\tau_2 = 27$ ns), while the decay time associated with recombination is $\tau_3 \sim 40$ ns \cite{Hogenbirk_2018}. Thus, LXe is, {\em a priori}, a brighter and (slightly) faster scintillator than LYSO, enabling improved energy resolution and time-of-flight performance, which in turn enhance the final image quality.  However, its attenuation length is almost three times longer than that of LYSO and the fraction of photons which interact via the photoelectric effect (relevant for PET application) a factor of two less.
  
 LXe has been used extensively in particle and astroparticle physics. In particular, the MEG experiment \cite{Gallucci_2009} deployed a large calorimeter
 ---800 litres volume filled with LXe---, with a thickness of 17 attenuation lengths and read out by 846 UV-sensitive photomultiplier tubes of 2-inch diameter. Other important examples are experiments searching for rare events, in particular neutrinoless double beta decay \cite{PhysRevLett.123.161802} and direct Dark Matter searches~\cite{Aprile:2010, xenon1T}. Future experiments such as nEXO \cite{Adhikari_2021} and DARWIN \cite{Aalbers_2016} are also based on LXe. All these detectors are Time Projection Chambers, able to read  the scintillation and ionization signals in LXe. Since the energy deposited by ionizing particles is distributed between both channels, the anti-correlation of the signals leads to a good energy resolution. For example, the XENON1T experiment measures 0.8\% (all resolutions are expressed in terms of rms through the paper, unless stated otherwise) at 2.46 MeV \cite{xenon1T} and the LZ experiment measures 0.67\% at 2.61 MeV \cite{LZ}.
 
The possibility of building  a PET scanner based on LXe was proposed more than four decades ago \cite{Lavoie} and explored by a number of subsequent works \cite{chepelRes, chepelEnergyRes, Doke1, miceli}.  The advantages of LXe compared to scintillating crystals are a large light yield, a fast decay time and the fact of being a liquid medium, which reduces the complexity of the system and increases its scalability. The disadvantage of having to add a cryogenic system to keep xenon in its liquid phase (the liquefaction temperature of xenon is $\sim$-112$^\circ$C at 1 bar) is not problematic, since  cryogenic systems reaching much lower temperatures are already in use in hospitals, for instance, to cool down magnets in magnetic resonance scanners. Two strategies were considered: a PET scanner based on scintillation only, and one that would combine both ionization and scintillation. Given the very high rate of PET scans, the second strategy may be unpractical, due to the long time (microseconds) needed to drift the charge. The LXe micro-PET prototype obtained an energy resolution of 4.3\% \cite{miceli} combining scintillation and charge, while the XEMIS1 prototype measured $5\%$ \cite{xemis_ER} with charge only. On the other hand, several groups tried prototypes based on pure scintillation (measured by PMTs), obtaining an energy resolution of 6.4--13.6\%  \cite{Doke1, miceli, ChepelEres}. None of these results were compelling enough to consider LXe as a viable alternative to standard inorganic crystals, in particular LYSO. Recently, the possibility of building a LXe-PET using modern SiPM technology has been considered by our group. SiPMs offer improved energy resolution and time-of-flight performance, compactness enabling one-to-one crystal coupling for enhanced interaction reconstruction, and insensitivity to magnetic fields, allowing PET-MRI compatibility, being a significant improvement over earlier PMT-based PET scanners. \cite{Gomez-Cadenas:2016mkq, Gomez-Cadenas:2017bfq, IEEE2018talk, Renner:2020ayj, Ferrario:2019pvw, Renner_2022, Nerea_2025}.

 \begin{figure}[htbp]
\centering 
\includegraphics[width=0.5\textwidth]{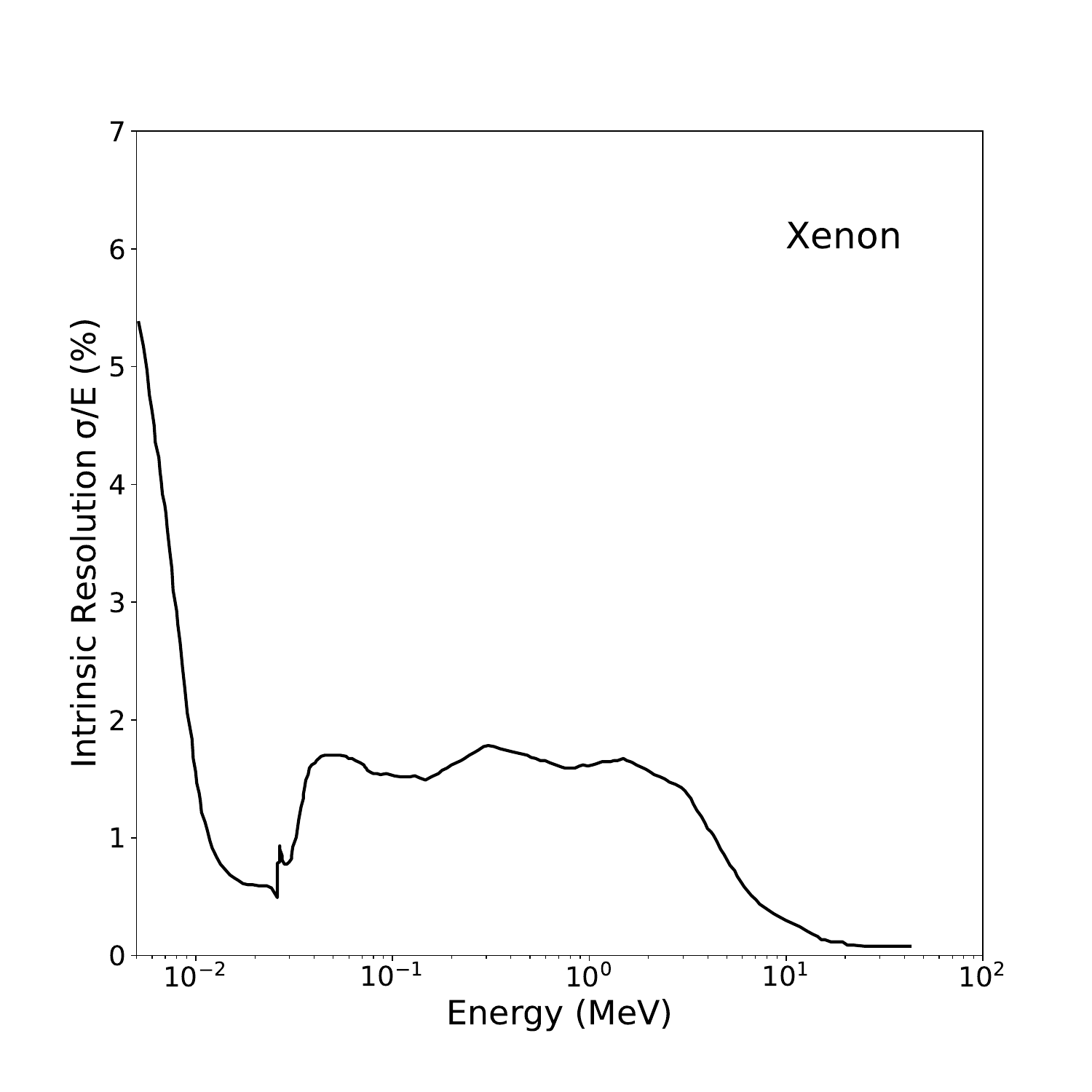}
\caption{\label{fig:dokei} Intrinsic resolution in LXe. Reproduced from Ref.~\cite{non-prop}. The original data were reported in $\%$ FWHM and have been converted here to $\%$ $\sigma$ (FWHM = 2.355 $\sigma$) for consistency with the rest of this work.}
\end{figure}

In order to assess the potential of LXe as calorimetric medium, in particular when using only scintillation light, it is essential to understand its 
intrinsic resolution ($R_i$). The intrinsic resolution of a scintillator refers to the fundamental limit at which Poisson statistics does not improve resolution due to underlying physics.
For example, the energy resolution of NaI(Tl), which, as mentioned above, is the most popular inorganic scintillator, is much worse than that expected from Poisson statistics in the energy region of MeV, thus partially spoiling its suitability for PET. The underlying physics explanation is the non-proportionality of the scintillation yield in NaI(Tl) for secondary electrons. 

In the case of LXe, the intrinsic resolution comes up due to two components: the non-proportionality of the scintillation light yield with energy and the fluctuations in electron-ion recombination due to scape electrons.  In 2002, Doke studied the non-proportionality of the scintillation light yield \cite{non-prop} resulting in $R_{i} \sim 1.8\%$ in the region between 100 keV and $\sim$5 MeV (Fig. \ref{fig:dokei}). His calculation is in good agreement with the modern Monte Carlo code describing the physics of LXe,  NEST \cite{nest}. Specifically, at 511 keV, NEST predicts $R_{i} \sim 1.3$\%. 

Several experimental groups have reported energy resolution measurements in LXe for different gamma-rays, but none of these obtained the intrinsic resolution component related to the non-proportional light yield. For instance, a work from Yamashita et al.  obtained an energy resolution of $\sim$12 $\%$  at 122 keV and 7.1 $\%$  at 662 keV \cite{YAMASHITA2004692}, while Aprile et al. reported 5$\%$ at 122 keV \cite{PhysRevC.84.045805}. The only attempt to go beyond simple energy resolution is that of Ni and collaborators, who estimated the intrinsic resolution as the combination of the non-proportionality of light yield and the electron-ion recombination fluctuations, obtaining 6-8 $\%$ \cite{Ni:2006zp} for gamma rays between 122 keV and 662 keV. Although this value is not directly comparable to the result obtained by Doke, this result is almost a factor $\sim$ 5 worse.

In this paper, we present a new measurement of the energy resolution in LXe, using scintillation only. A value of the intrinsic resolution (non-proportionality + recombination fluctuations)  is also shown. Our setup is conceptually identical to the SSB described above, and illustrated in Fig. \ref{fig:ssb}, where the scintillating material is now LXe. The advantage of the SSB is its large light collection efficiency, as well as the availability of VUV-SiPMs with large PDE. This results in a large number of detected photoelectrons (pe), thus minimizing Poisson fluctuations and leading to a measurement  close to $R_{i}$. 

On the other hand, we have seen that SSBs based on a variety of inorganic scintillators constitute the building blocks of calorimeters that have been used or are planned for future experiments in nuclear and particle physics, as well as the basic building block for PETs. The obvious question, then, is whether an SSB based on LXe  could be a viable alternative for some of the future applications, and in particular for LYSO blocks in PET applications.

\section{SETUP}
\begin{figure}[htbp]
\centering 
\includegraphics[width=0.5\textwidth]{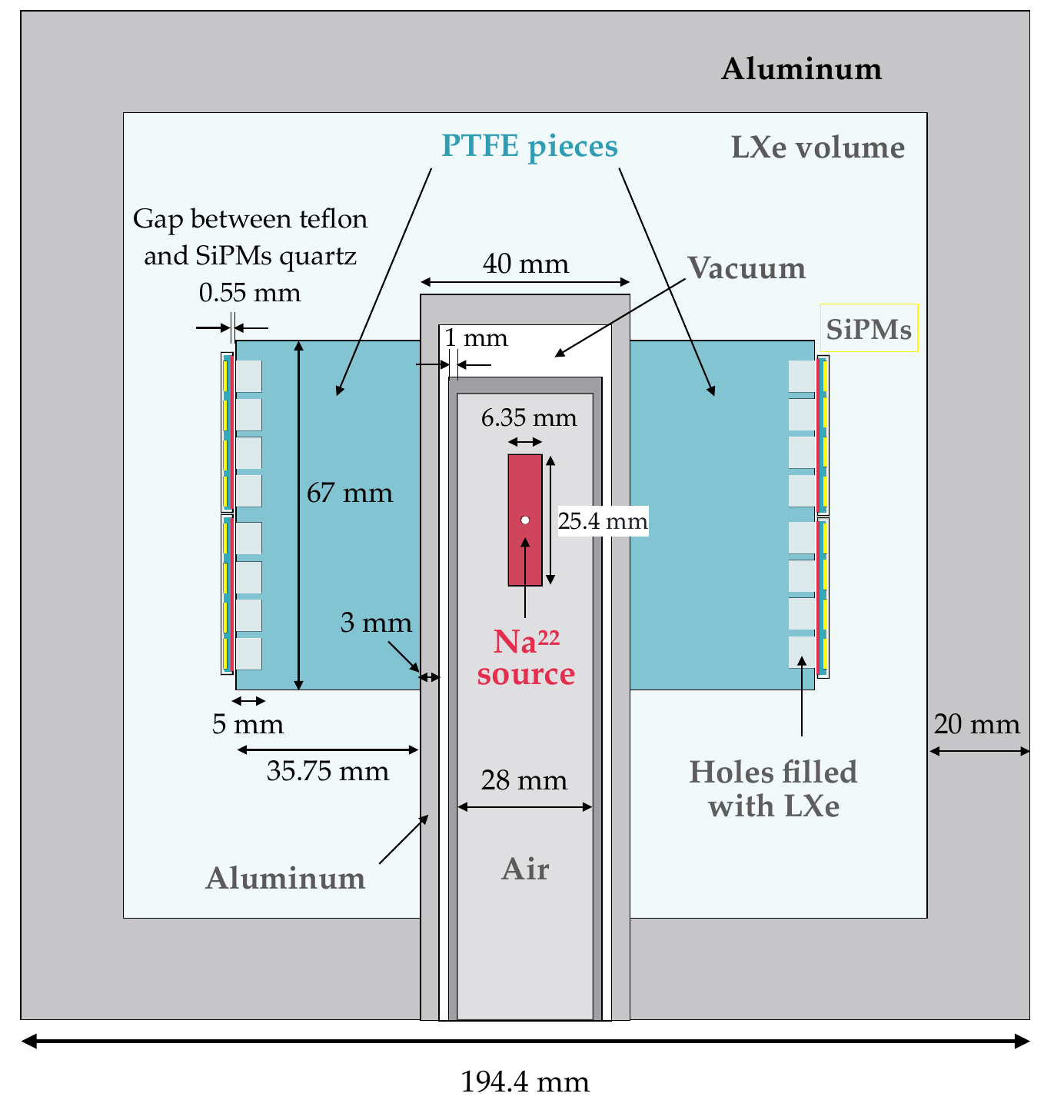}
\includegraphics[width=0.2\textwidth]{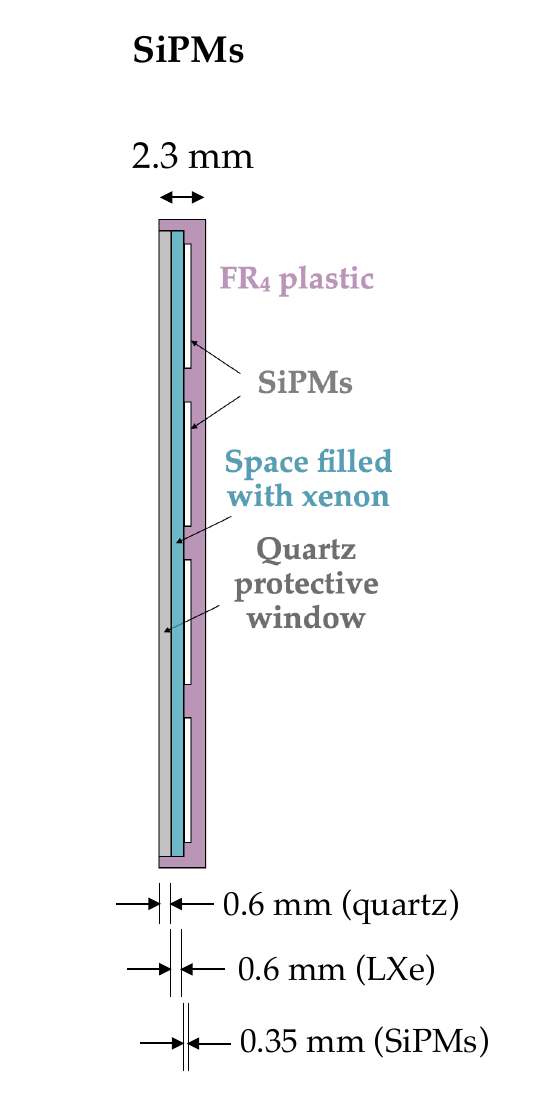}
\caption{\label{fig:setup} A diagram of our experimental setup (top), with an enlarged view of the SiPM arrays (bottom). The dimensions of the relevant parts of the system are specified.
}
\end{figure}

Our setup, shown schematically in Fig. \ref{fig:setup}, consists, conceptually, of two LXe SSBs. A $\ensuremath{^{22}}$Na calibration source, located between the two SSBs, provides pairs of almost back-to-back gammas with an energy of 511 keV. Some of the gammas interact in the xenon volume, producing VUV light, which propagates inside the channel in which the interaction occurs. A large fraction of this light reaches the SiPM that (almost) closes the channel. Each SiPM, therefore, provides an independent measurement of the $\ensuremath{^{22}}$Na spectrum. Fitting the region close to the photopeak provides a measurement of the energy resolution, $R_{m}$.

\begin{figure}[htbp]
\centering 
\includegraphics[width=0.5\textwidth]{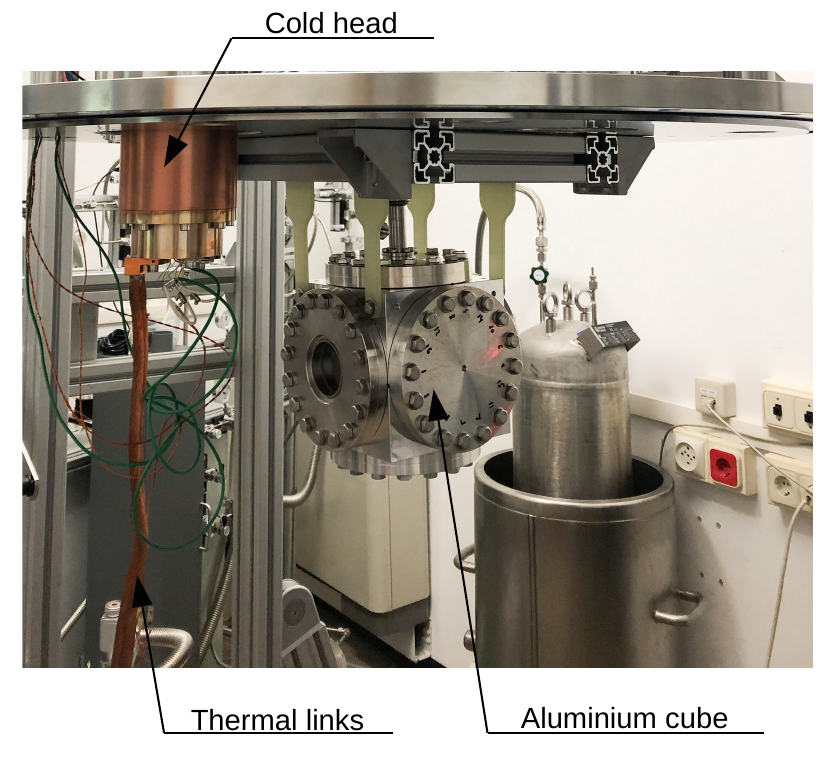}
\caption{\label{fig:cube} The aluminum cube holding the two SSBs used for the measurement. Notice the thermal links attached to the cold head.}
\end{figure}

The active region of the setup consists of two LXe SSBs housed in a metal-sealed aluminum cube (F100-1ALU) from Vacom. The xenon is continuously recirculated in gas phase by a double diaphragm compressor and flows through a model PS4 MT15 R2 hot getter from Sigma Technologies. This process removes nitrogen, oxygen and water from the xenon, all of which quench xenon scintillation light \cite{impurities}. The internal cube is housed within a large vacuum vessel so that the structure is thermally isolated while cooled. Said cooling is performed using a Sumitomo CH-110 cold head coupled to the cube via custom-made copper thermal links. Heat dissipation from the thermal links is minimized by wrapping them in polyethylene terephthalate foil. The cube and cooling machinery are visible in Fig. \ref{fig:cube}. The gas passes through a heat exchanger when entering and exiting the inner cube, which precools the gas before liquefaction for operation at 1.26 bar.

\begin{figure}[htbp]
\centering 
\includegraphics[width=0.5\textwidth]{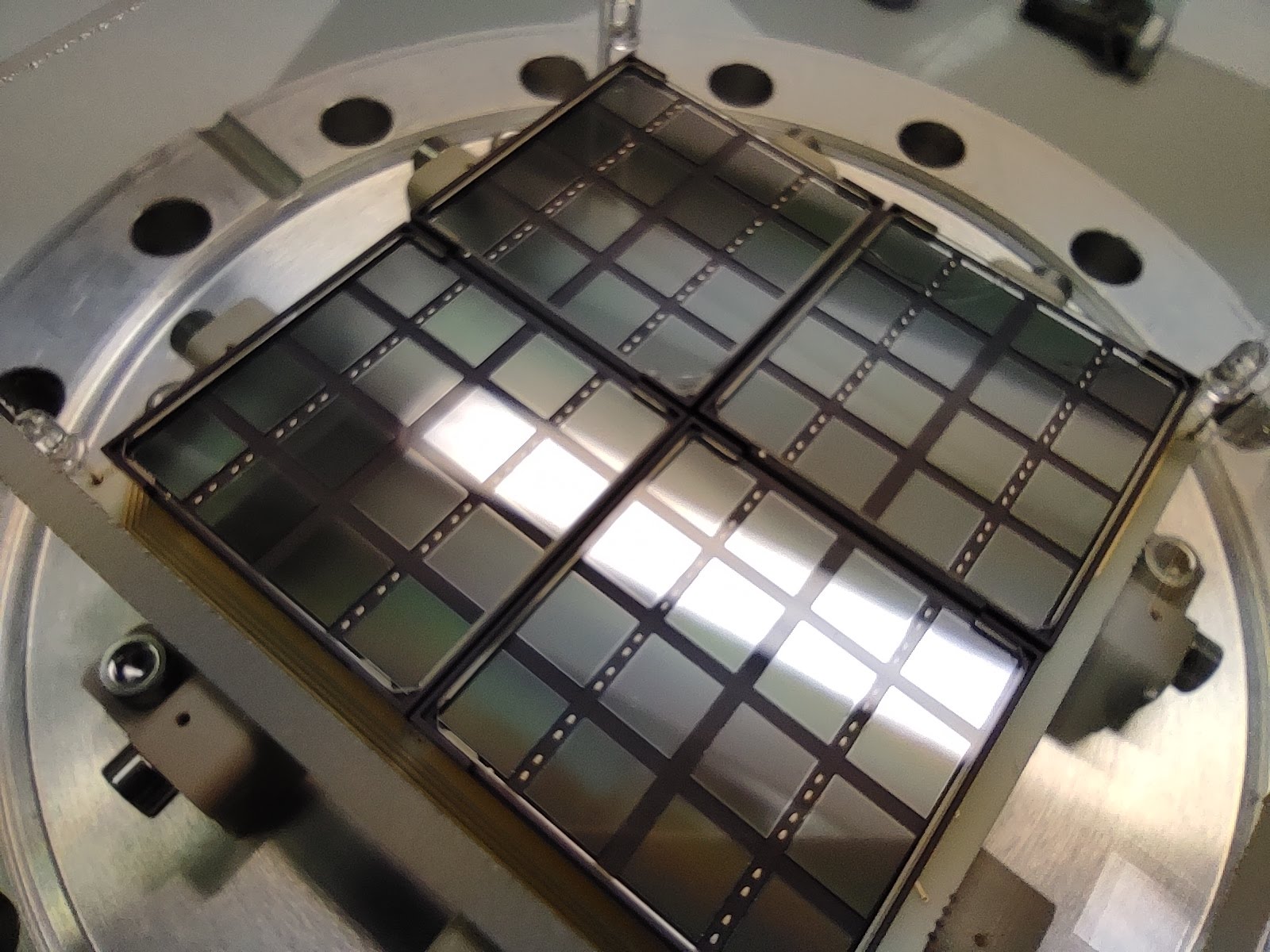}
\includegraphics[width=0.5\textwidth]{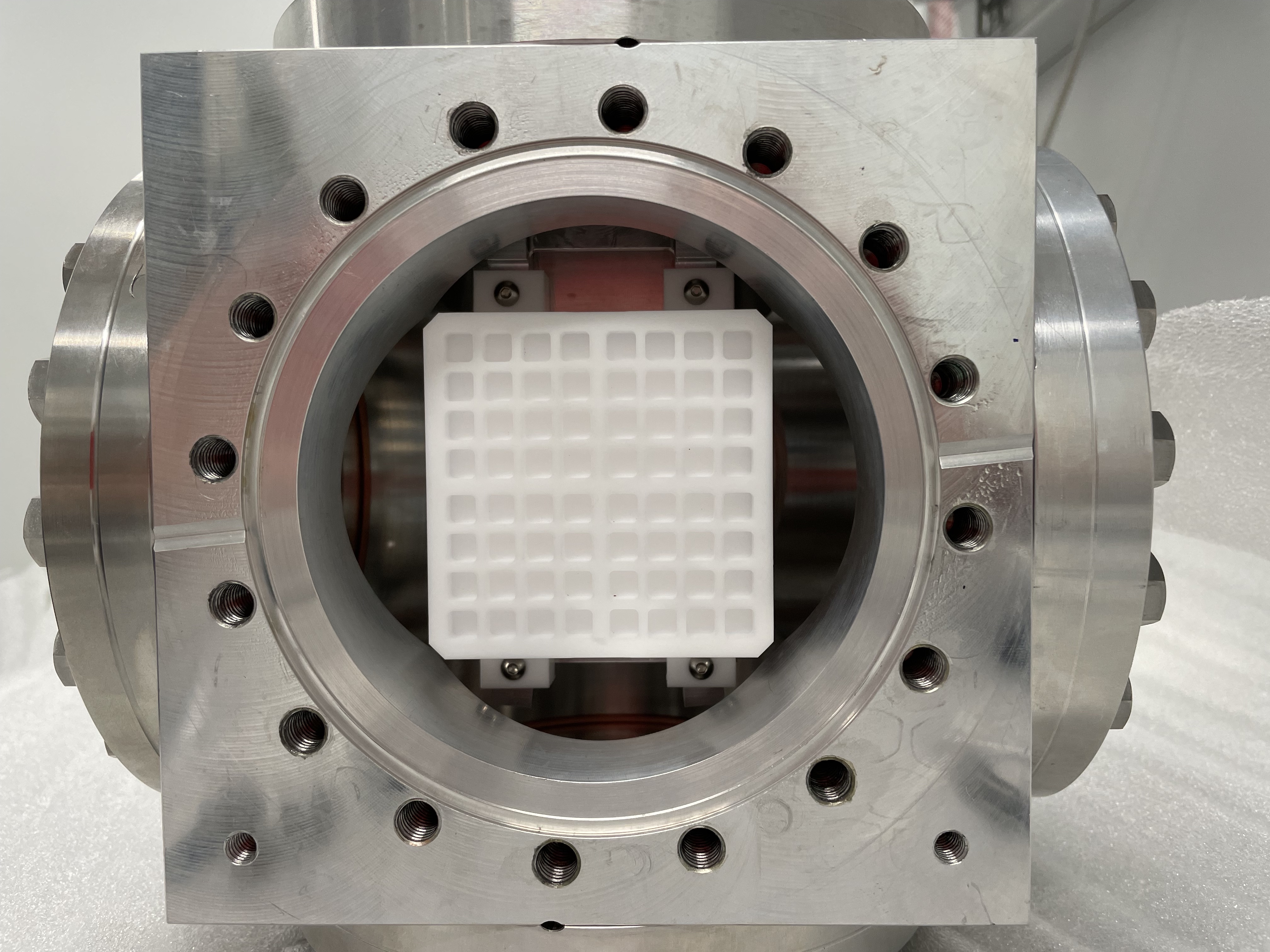}
\caption{\label{fig:sns} Top: instrumented plane of four arrays of 4$\times$4 SiPMs. Bottom: teflon piece defining the SSB.}
\end{figure}
Fig. \ref{fig:sns} shows the main components of the SSB before mounting and filling with xenon. Fig. \ref{fig:sns}-top shows an array of Hamamatsu S15779(ES1), VUV-sensitive SiPMs. They have 30\% photodetection efficiency at the peak wavelength of xenon scintillation and are protected by a quartz window with $\geqslant$~90\% transmission at relevant wavelengths. Their active area is $5.95\times5.85$ mm$\ensuremath{^{2}}$ and they are packed in arrays of 4$\times$4 sensors. Each SiPM reads out a 6$\times$6 mm$^2$, 5-mm deep hole in a polytetrafluoroethylene (PTFE) piece (Fig. \ref{fig:sns}-bottom). The two SSBs are mounted inside the aluminum cube with the sensitive face of the SiPMs facing each other.
The teflon pieces are mounted to leave a 0.5 mm gap between the SiPM face and the end of the empty cell. In this way the LXe can easily fill the gaps, completing the SSB cells. The result is that each SiPM is exposed to the scintillation light of gamma interactions occurring only in the hole in front of it. The high reflectivity of the PTFE at VUV wavelengths ($\sim$98\%)\cite{AKERIB201334, YAMASHITA2004692} ensures a high geometric light collection efficiency.

Signals are digitized using two TOFPET2 ASICs \cite{petsys, IEEE2018talk}, one for each read-out plane, mounted in vacuum on the outside walls of the cube corresponding to the relevant SiPM array. They integrate the detected charge and provide a fast timestamp close to origin to reduce signal degradation before digitization.

Welded to the bottom flange of the cube is a rectangular aluminum calibration port, sealed at the top end and extending above the center of the cube. This allows the insertion of sources between the two SSBs for coincidence measurements using back-to-back gammas. For the main measurement reported here, a $\ensuremath{^{22}}$Na calibration source is placed in the center of the cube inside the calibration port. The radioactive material (with an activity of $\sim260$ kBq is a sphere of 0.25 mm diameter, which can be considered point-like, and is encapsulated in a plastic support. Other two calibration sources are also used, with similar encapsulation: a $\ensuremath{^{133}}$Ba source with $\sim700$ kBq of activity and a $\ensuremath{^{57}}$Co source of $\sim40$ kBq of activity.

\section{MONTE CARLO SIMULATION}
\label{sec:mc}
A detailed simulation of our set-up has been developed using a Monte Carlo software application based on \textsc{Geant4} \cite{Geant}. The physical and optical properties of the materials in use are simulated thoroughly; in particular, the reflectivity of PTFE has been modeled as a Lambertian reflectivity with a value of $98\%$ and the value of the refractive index as a function of the wavelength has been extracted from Ref.~\cite{Baldini_2006}. The other physical parameters relevant for optical photon propagation are the Rayleigh scattering length (36 cm \cite{RayleighLXe}) and the transparency and refractive index of the quartz window placed in front of the SiPMs to protect them (90$\%$ and 1.6, respectively). The photodetection efficiency provided by the sensor datasheet is also included as a function of the wavelength of the optical photons. We turn off any intrinsic fluctuations of the LXe response to gamma interactions, so that the amount of scintillation photons produced is proportional to the deposited energy (see more detailed information in Appendix \ref{append}). 

\begin{figure}[htbp]
\centering 
\includegraphics[width=0.5\textwidth]{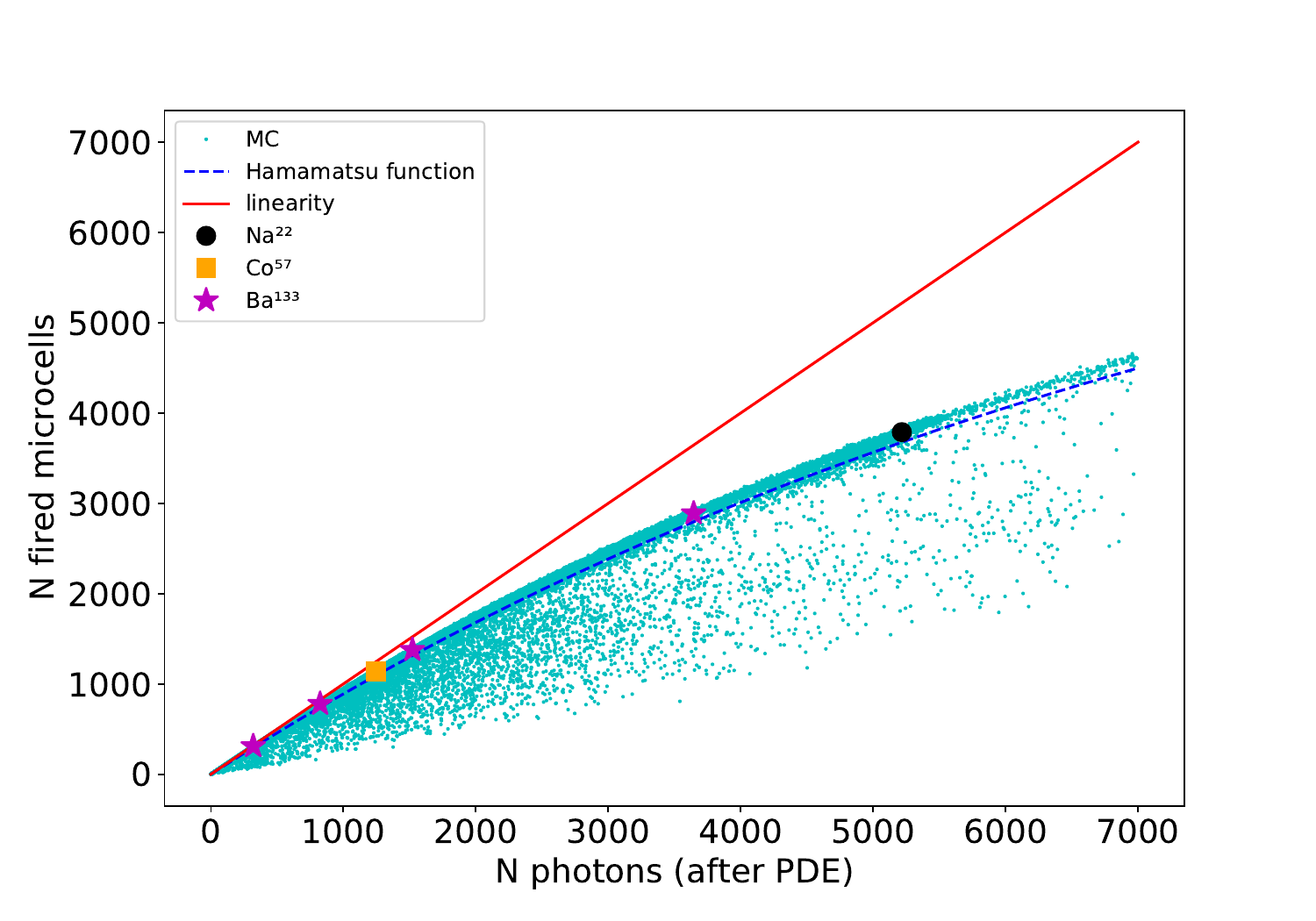}
\caption{\label{fig:satcurve} Number of fired microcells versus the actual number of photons that should be detected (after the photodetection efficiency) in our simulation (blue points). The positions of the photoelectric peaks of three calibration sources with and without saturation effects are highlighted, as well as the saturation curve provided by Hamamatsu. The perfect linearity case is drawn in red to guide the eye.}
\end{figure}

The simulation of a $\ensuremath{^{22}}$Na source placed in the calibration port provides a charge distribution with the photopeak centred in 5240 pe. Since the number of microcells in our SiPMs is 6162, our measurement is affected by saturation. In order to confirm that, we ran a second simulation, where each microcell of the SiPMs is simulated and the recovery time of our sensors (158 ns at $\sim$ -110$^\circ$C) taken into account, to model the loss of photons due to saturation;  we found that in this case the photopeak is centered in 3752 pe, which implies a significant charge loss. The same simulations, with and without saturation effects, were carried out for two other available calibration sources, with lower energy gammas, namely $\ensuremath{^{133}}$Ba (provides gamma lines at 30.8, 81 and 356 keV) and $\ensuremath{^{57}}$Co (provides gamma line at 122 keV). In Fig. \ref{fig:satcurve} the number of fired microcells versus the number of photons that should be detected in the case of negligible saturation  is shown for MC events at all energies. The saturation curve provided by Hamamatsu \cite{hama_curve} is superimposed on top of this distribution, showing the excellent match with the simulation. The positions of the photopeaks of the three calibration sources lie close to the Hamamatsu model.

Therefore, in the following we will use the Hamamatsu saturation curve to correct data for saturation, in order to provide a reliable measurement of the energy resolution of the $\ensuremath{^{22}}$Na source and for the gamma of 356 keV of the $\ensuremath{^{133}}$Ba source. The low-energy peaks do not need to be corrected with this curve as they are not affected by the SiPM saturation.

\section{MEASUREMENT} \label{meas}

During a run, a timestamp is provided by the ASIC whenever the recorded charge in a channel exceeds a lower threshold. If the charge exceeds a second, higher threshold, the ASIC starts integrating it until it goes below this threshold again. If the second threshold is not reached, the integrated charge is not saved. 

The data acquisition output is a list of channels with integrated charge and corresponding timestamps. Once the channels are divided event by event, a first filter is applied to retain only coincidences, i.e., events where at least one sensor from each detection plane is present. This condition is applied only to data taken with the $\ensuremath{^{22}}$Na source.
Given the cell structure of the SSBs, most of the charge is read out by the sensor coupled to the xenon volume where the interaction occurred, therefore,  for all sources we single out the sensor with the larger detected charge for each plane. 

\begin{figure}[htbp]
\centering 
\includegraphics[width=0.50\textwidth]{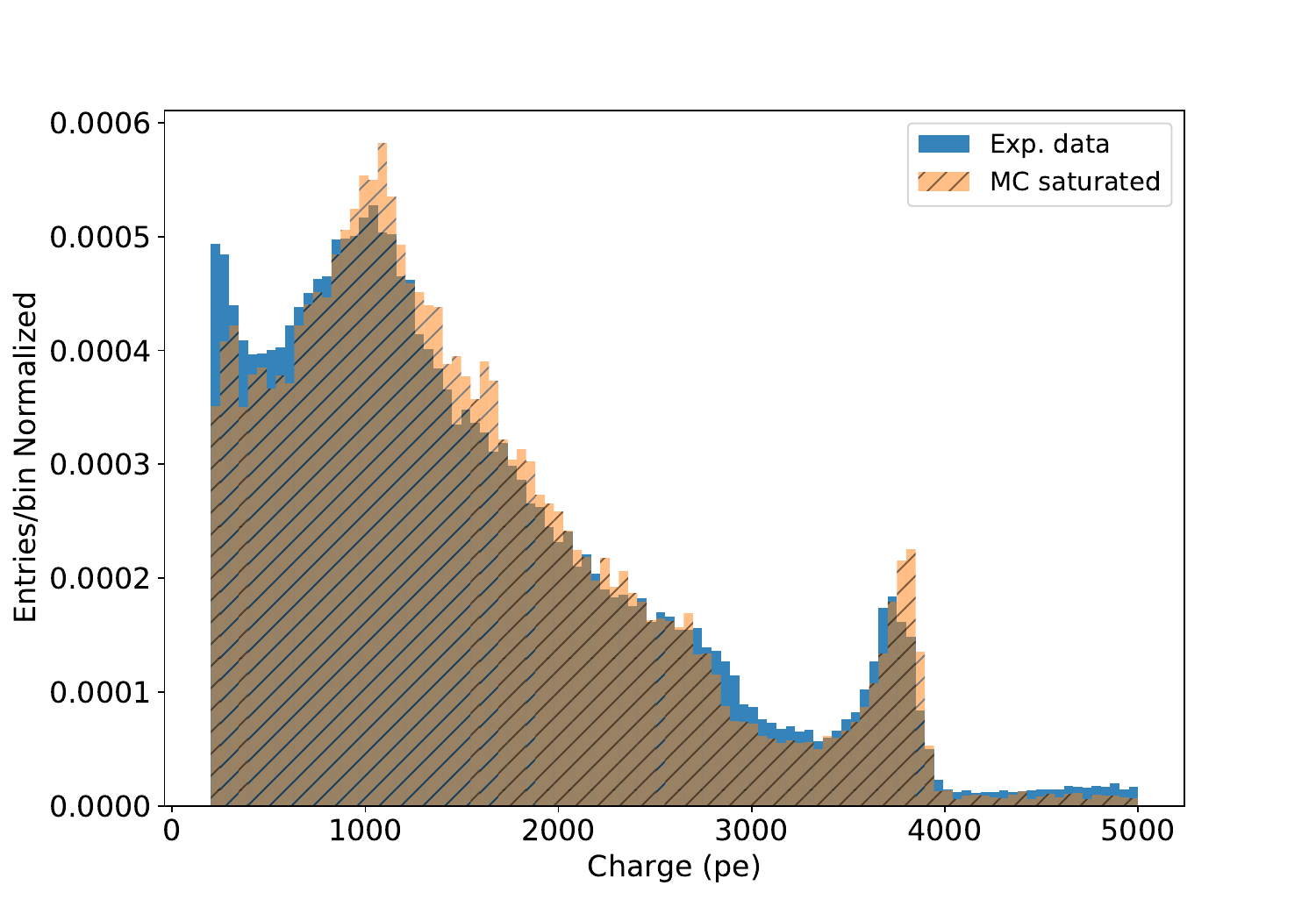}
\caption{\label{fig:channel} Charge distribution detected by one of the SiPMs without saturation correction and the charge distribution obtained in the MC saturated data for the $\ensuremath{^{22}}$Na source.}
\end{figure}

Once the energy spectrum was obtained for the three sources, the ASIC non-perfectly linear response with energy due to the integrator was corrected and the energy spectrum in pe was obtained (see Fig. \ref{fig:channel}).

\begin{figure}[htbp]
\centering 
\includegraphics[width=0.5\textwidth]{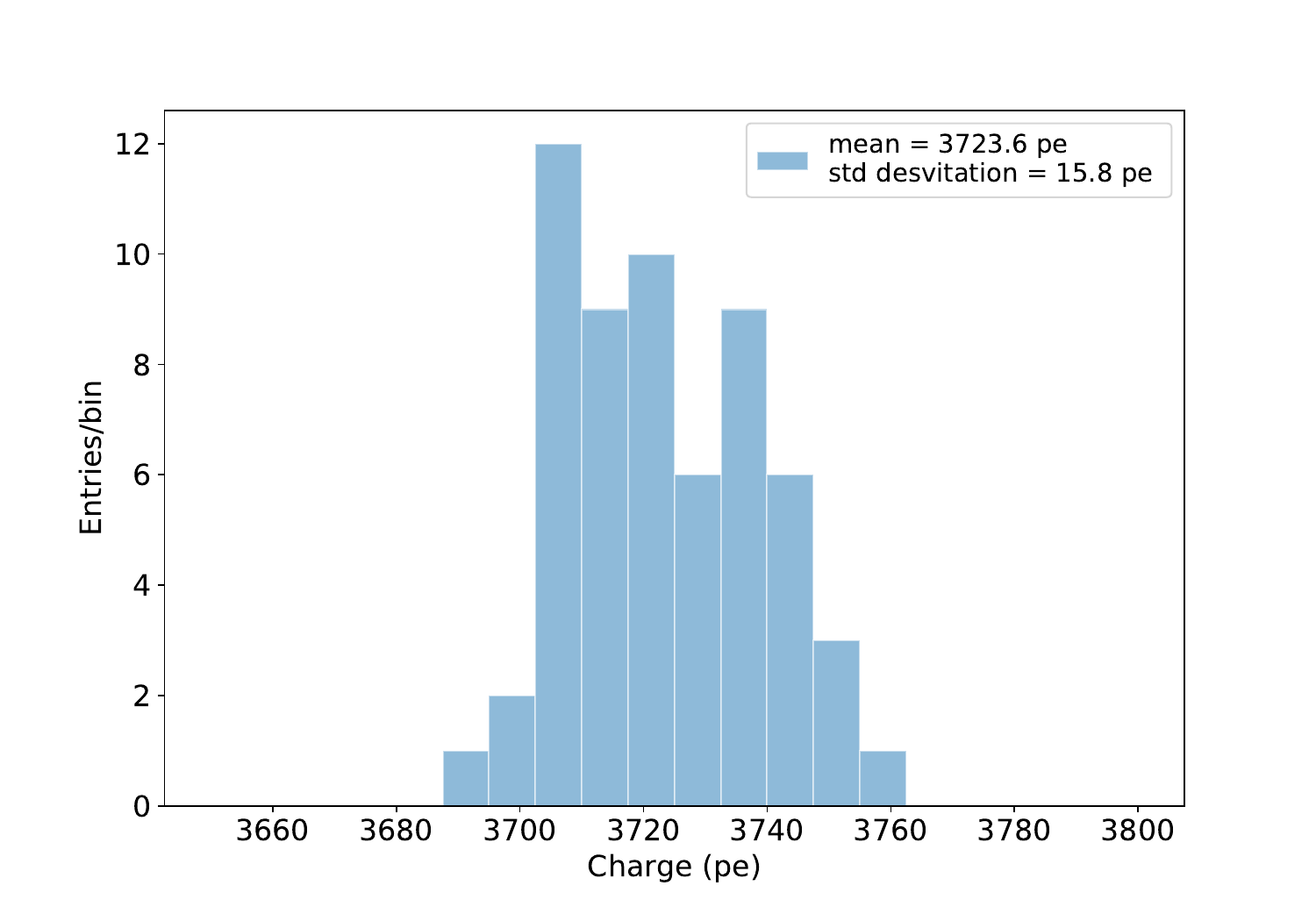}
\caption{\label{fig:mean} Distribution of the raw peak position in each cell in the reference plane for the $\ensuremath{^{22}}$Na source. 
}
\end{figure}

Fig. \ref{fig:channel} shows the photoelectric peak together with the Compton shoulder for one channel, showing good agreement between the experimental data and MC simulated saturated data. Fig. \ref{fig:mean} shows the distribution of the peak positions for the SiPMs in the reference SSB used in the measurement. The reason for choosing only one SSB is that one SSB is aligned better than the other and the response of the different sensors is more uniform. The distribution shows a spread of around 0.4$\%$. Instead of equalizing the sensor responses, we consider the resolution in each of them as an independent measurement and compute a weighted mean using the following procedure. A normal distribution is fitted to a region around the peak to extract the resolution. To choose the region, we gradually extend the region on both sides of the peak until the fit quality —quantified by the $\chi^2$ value— starts to deteriorate, indicating that the data is no longer well described by a Gaussian distribution. All resolution values (defined in Eq. \ref{eq:res}) for accepted fit regions are kept, and their mean and fit error are taken as the representative resolution.

\begin{equation}
R = \frac{\sigma}{\mu}\,
\label{eq:res}
\end{equation}

where $\sigma$ and $\mu$ are the best values for the Gaussian parameters coming from the fit. This procedure is applied to all the sensors in the SSB and their results are combined using a weighted mean, where each value is weighted by the inverse of its squared fit error (i.e., more precise measurements have a larger influence on the final resolution), thus yielding a final resolution of $2.5 \pm 0.4$ $\%$  at 511 keV. The total uncertainty is calculated as the combination of two contributions: the statistical uncertainty from the fits and the spread across sensor resolutions.

\begin{figure}[htbp]
\centering 
\includegraphics[width=0.5\textwidth]{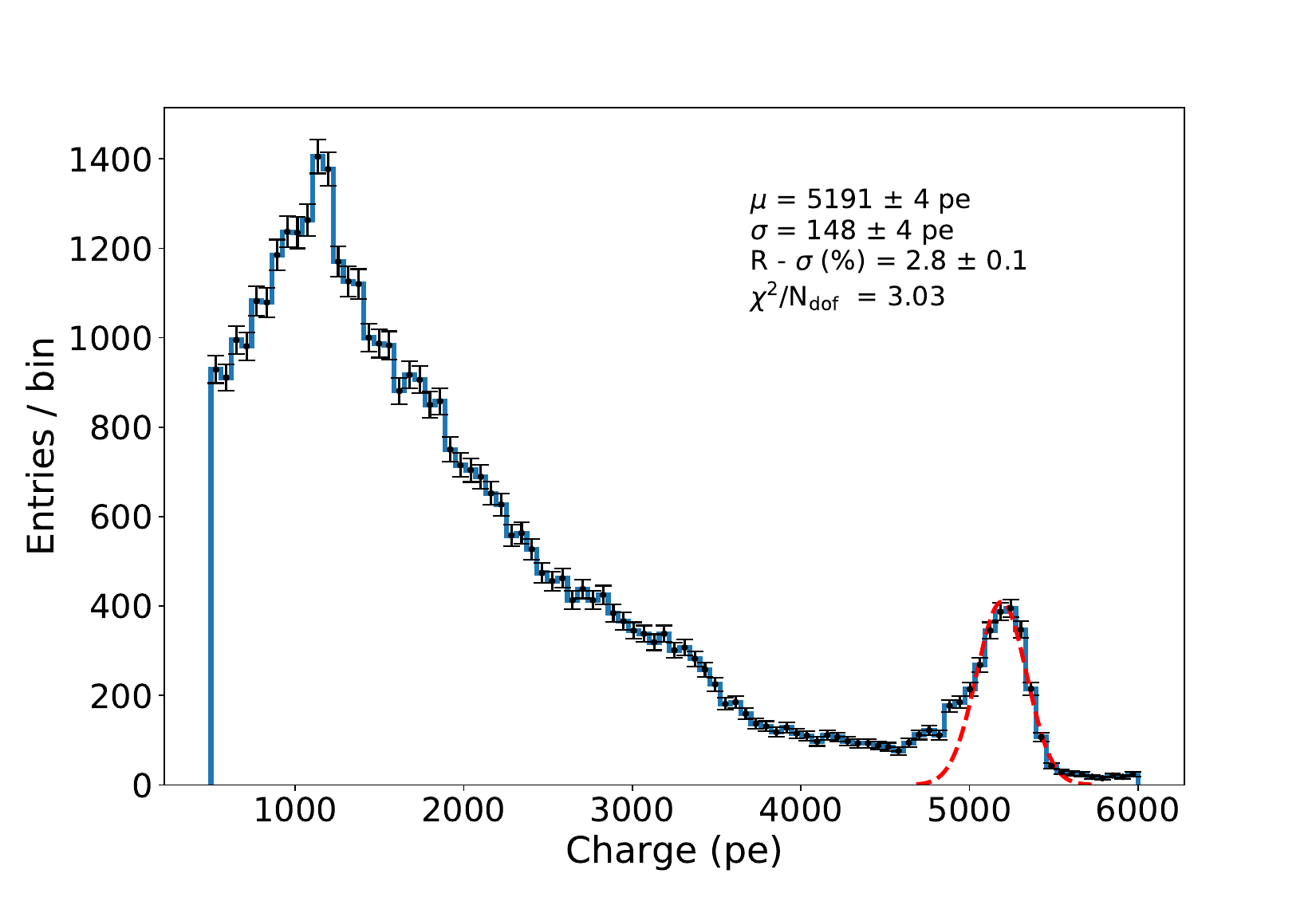}
\caption{\label{fig:MC} Charge distribution obtained in the MC without SiPM saturation with a gaussian fit in the $\ensuremath{^{22}}$Na peak. 
}
\end{figure}

The energy resolution is slightly better than the value predicted by MC without SiPM saturation, which is $2.8 \pm 0.4$ $\%$  at 511 keV (see Fig. \ref{fig:MC}). The MC error includes the uncertainties in the optical modelling, such as the reflectivity of the teflon and the reflection and refraction between LXe and the quartz window that protects the SiPMs (see Fig.~\ref{fig:setup}\textit{--bottom}). 

\begin{figure}[htbp]
\centering 
\includegraphics[width=0.5\textwidth]{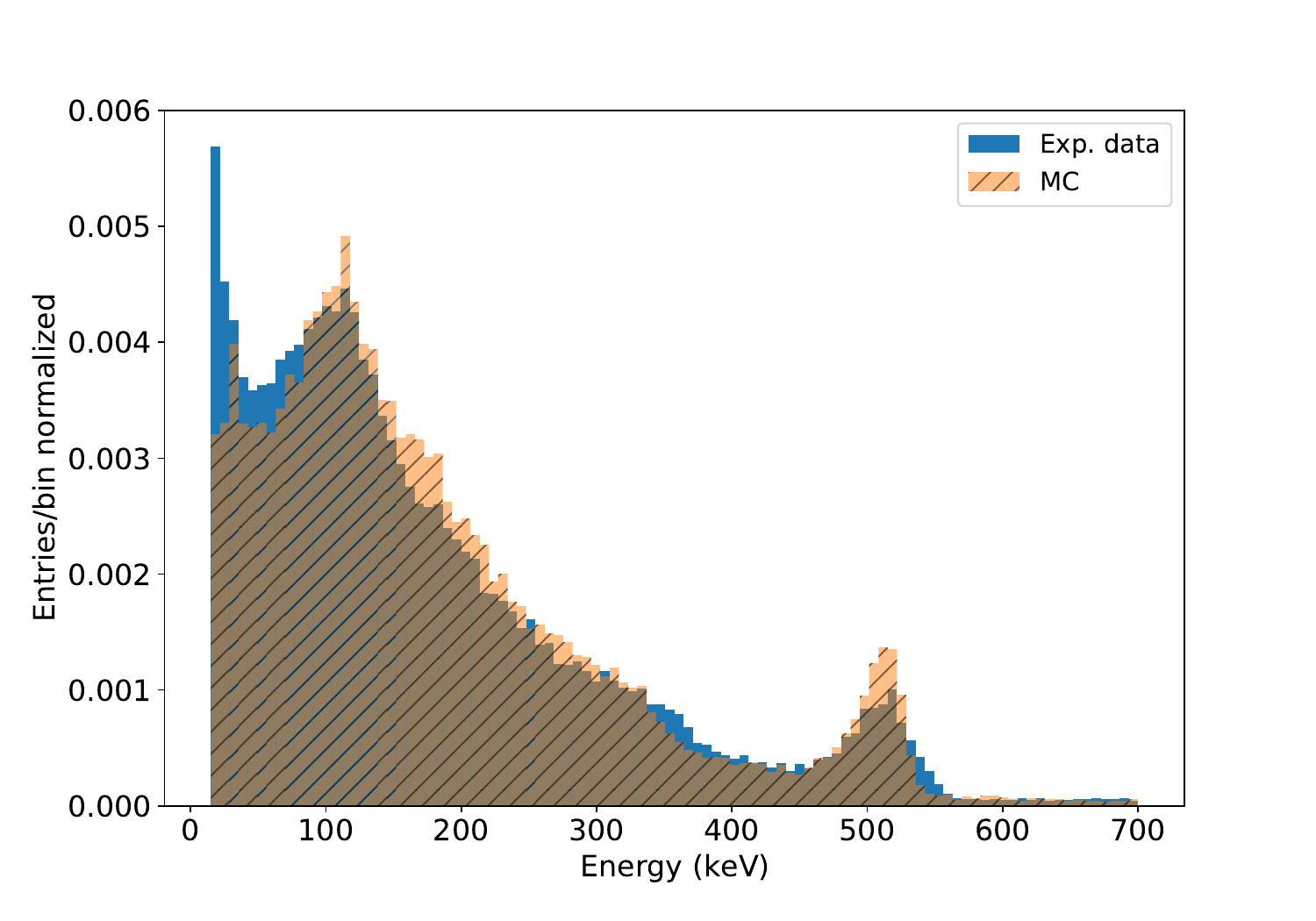}
\caption{\label{fig:result} Energy distribution by one of the SiPMs after saturation correction and  the energy distribution obtained in the MC for the $\ensuremath{^{22}}$Na source. 
}
\end{figure}

\begin{figure}[htbp]
\centering 
\includegraphics[width=0.5\textwidth]{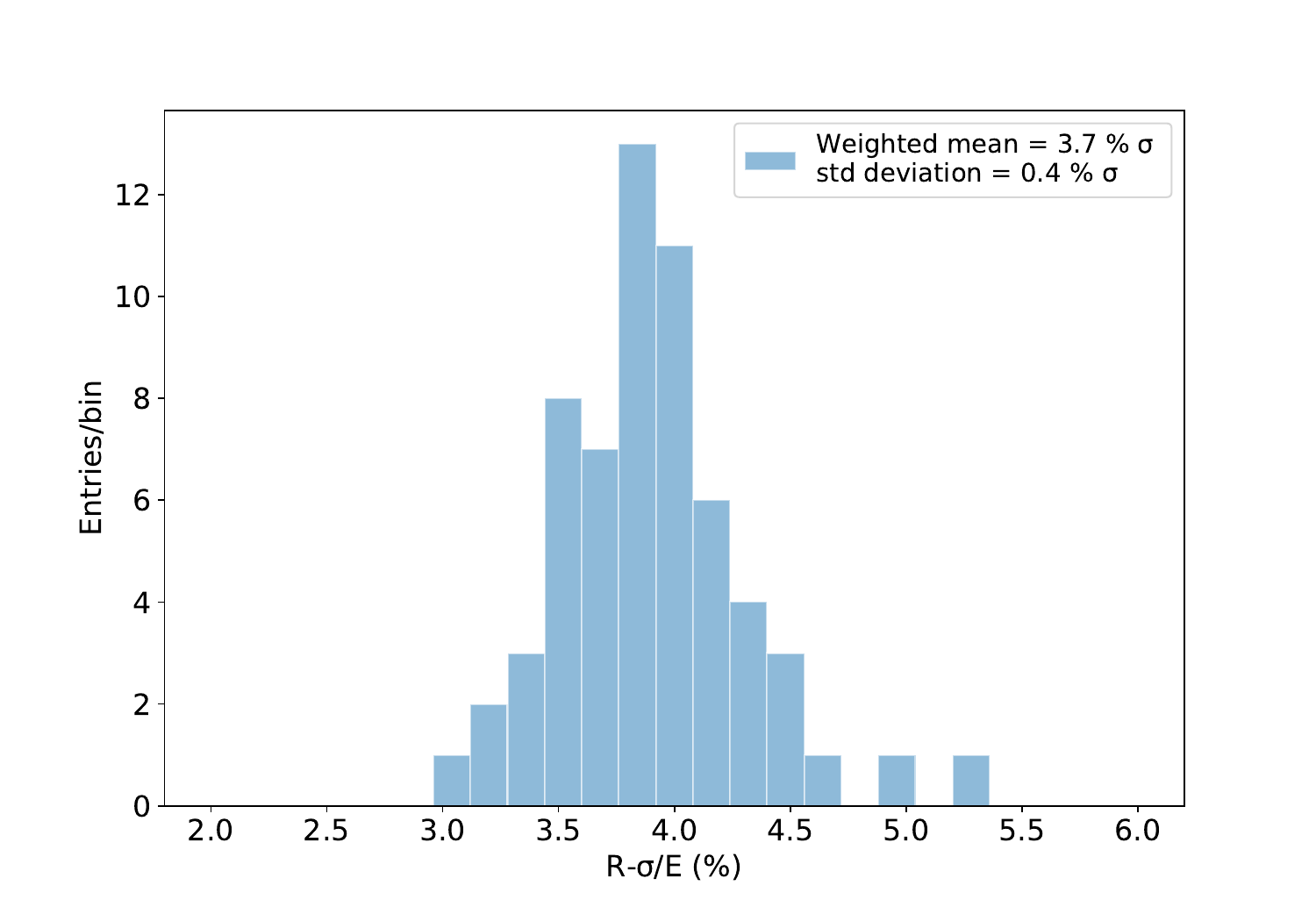}
\caption{\label{fig:result_all} Energy resolution after saturation correction for all SiPMS in the reference SSB for the $\ensuremath{^{22}}$Na source. }

\end{figure}

The low value of the energy resolution compared to MC is due to SiPM saturation, which has the effect of compressing the data and brings on an artificial better energy resolution. Therefore, correcting the data with the saturation curve provided by Hamamatsu, and fitting the peaks with the same procedure as before, an energy resolution of $3.7 \pm 0.4$ $\%$ at 511 keV is obtained, showing compatibility with the MC result within  $1.5\sigma$.  Fig. \ref{fig:result} shows the overlaid energy spectra of the experimental and MC data without saturation showing a good agreement. Fig. \ref{fig:result_all} presents the corrected energy resolutions for all the SiPMs in the reference SSB. 

\section{RESULTS}
Following the same procedure for the three sources used, we obtain the energy resolution for 5 gamma peaks. 

\begin{figure}[htbp]
\centering 
\includegraphics[width=0.5\textwidth]{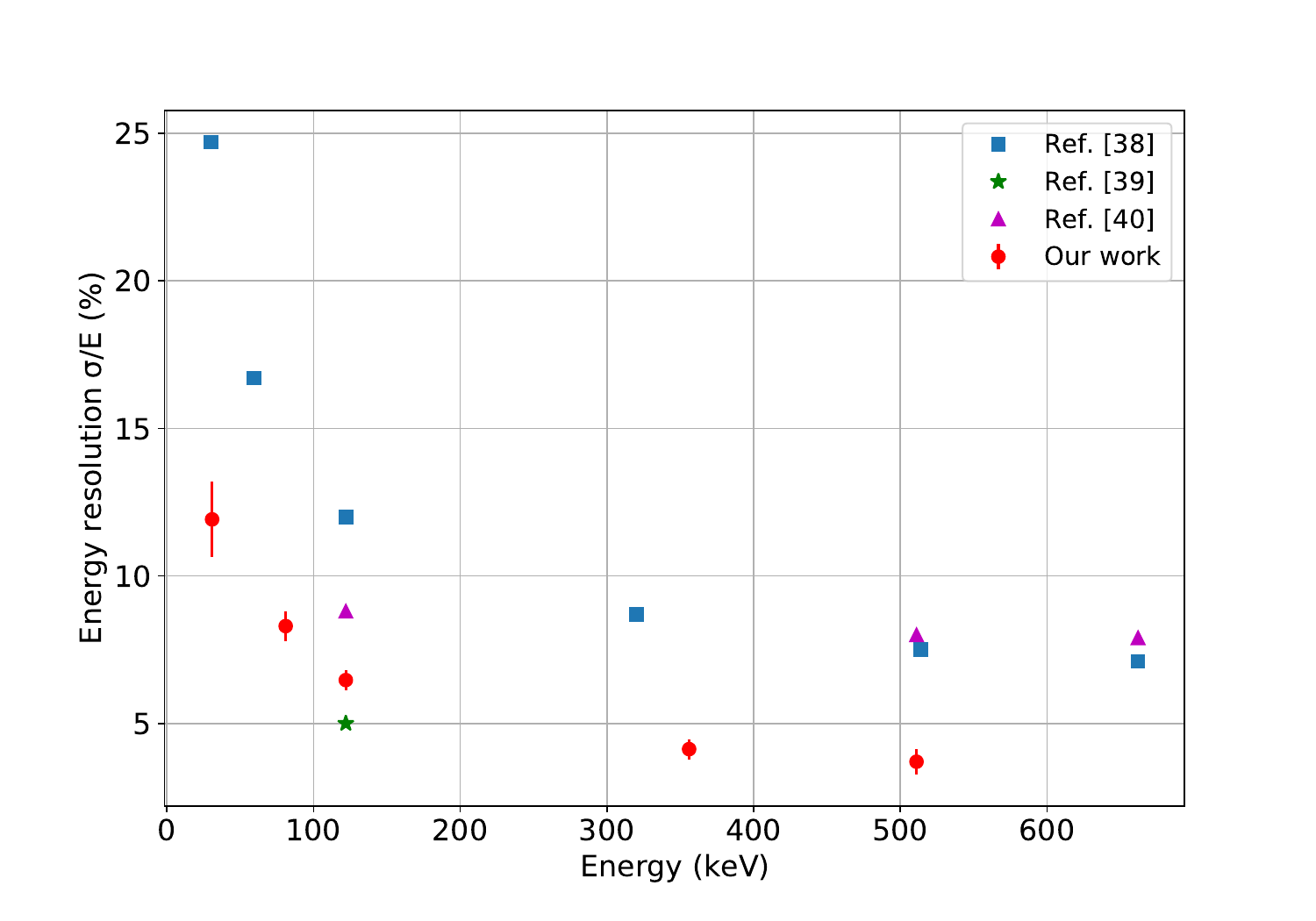}
\caption{\label{fig:energy_res} Energy resolution for different gamma rays in LXe. }

\end{figure}

The energy resolution obtained in our work is shown in Fig.\ref{fig:energy_res}, together with others' measurements. It can be seen that our results are 2-3 times better than previous works using only scintillation. The only slightly better measurement is the one from Ref. \cite{PhysRevC.84.045805}, which only measured one peak. 

Following Doke  \cite{non-prop}, the measured resolution, $R_m$, can be seen as the quadratic sum of the the resolution related to the non-proportionality of scintillation light yield with energy, $R_i$, the resolution due to Poisson variation in photon detection, $R_p$ and the resolution coming from the fluctuations of electron-ion recombination due to escape electrons, $R_r$:

\begin{equation}
R_m^2 = R_i^2 + R_p^2 + R_r^2.
\label{eq:intr}
\end{equation}

$R_p^2$ comes from the variations of light collection due to the detector geometry and the statistical fluctuation of number of photoelectrons from the SiPMs (the reflections on PTFE and on the quartz windows in front of the SiPMs, the geometric acceptance, the sensor PDE and the amplification stages of the electronics). 

Since our MC simulation does not include intrinsic fluctuations, we extract the intrinsic resolution (non-proportionality and fluctuations due to recombination), which we call $R_c$, by subtracting the Poissonian width as determined by MC from the measured resolution.

\begin{figure}[htbp]
\centering 
\includegraphics[width=0.5\textwidth]{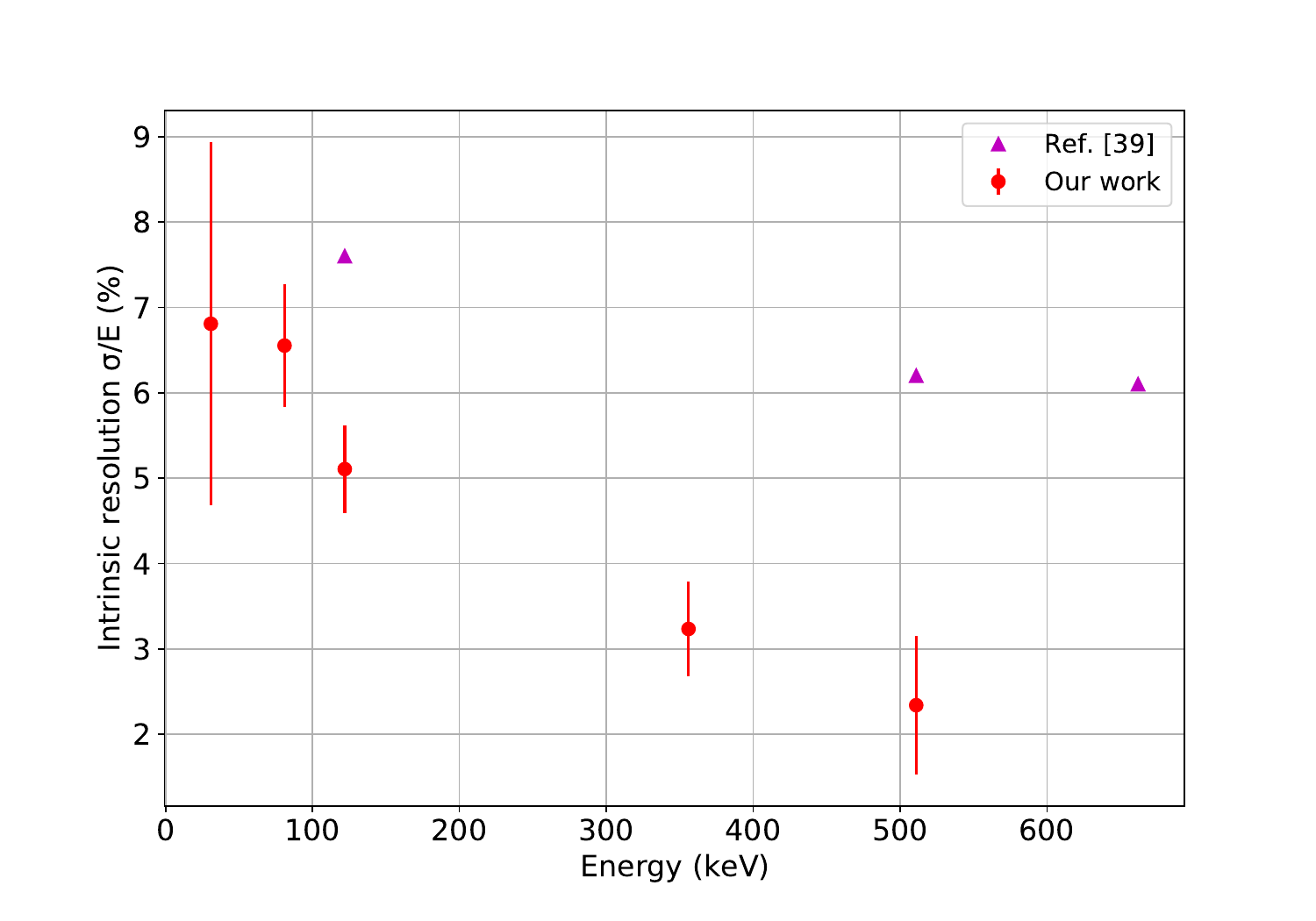}
\caption{\label{fig:intr} Intrinsic resolution calculated following Eq.} \ref{eq:intr} for the three calibration sources.
\end{figure}

Using Eq.~\ref{eq:intr}, we find $R_c = 2.3 \pm 0.8$ $\%$  at 511 keV, compatible with Doke's \cite{non-prop} $\sim 2$ $\%$, taking into account that our value includes also the recombination fluctuation. In Fig.~\ref{fig:intr} the values of the intrinsic resolution are shown for the three calibration sources used in the set-up. Opposed to Ref.~\cite{non-prop}, our results seem to indicate that its value changes significantly in the 100-1000 keV range.

\section{DISCUSSION}

In this work, we have measured the energy resolution attainable in a segmented scintillation block (SSB) filled with LXe and read out by VUV-SiPMs. The SSB was designed to maximize light acceptance and the SiPMs employed have the largest PDE currently available in the market. For 511 keV, an energy resolution of $3.7 \pm 0.4$ $\%$ was obtained, a factor of two better than the result reported in Ref. \cite{YAMASHITA2004692}.

The intrinsic resolution  $R_c^2$ that our procedure yields at 511 keV is compatible with that of Doke's \cite{non-prop}. However, it increases as the energy lowers. Previous measurements of this intrinsic resolution yielded very high values (6-8 $\%$) \cite{Ni:2006zp}, not compatible with our current measurement.

Furthermore, our work opens new possibilities for apparatus based on liquid xenon scintillators using scintillation only. In particular, it suggests that SSBs using LXe as a scintillator can be very competitive as building blocks in segmented calorimeters, with applications to nuclear and particle physics as well as PET technology. While there is ample room for technological development eventually increasing the PDE of the SiPMs), our measurement clearly suggests that a LXe-SSB can compete and eventually outperform, in terms of energy resolution, those using other scintillators.

\appendix
\section{Details of the Monte Carlo optical simulation}
\label{append}

A dedicated Monte Carlo simulation of the experimental setup has been developed using the \textsc{Geant4} toolkit \cite{Geant}.  This appendix provides additional details on the optical modeling assumptions and on the evaluation of the systematic uncertainties associated with the optical transport of scintillation photons in liquid xenon.
\newline
\newline
\textbf{Optical surface model}
\newline

The optical properties of the detector materials were implemented using the \textsc{Geant4} unified model for optical photon interactions at surfaces. In this framework, the reflection properties of the PTFE were modeled assuming a purely diffuse (Lambertian) reflection component with a reflectivity of 98$\%$.

The choice of a Lambertian reflection model is motivated by several measurements reported in the literature. In particular, the study by Silva et al. \cite{Silva2010} mentions more than 90$\%$ of the reflection is due to diffuse lobe at 0 degrees of incident angle, and more than 80$\%$ at 65 degrees. 

Other studies have reported the presence of a specular component that increases for large angles of incidence (above $\sim 60$–70°) \cite{EpjcPTFE}. However, the authors of that work also note a disagreement with previous results \cite{Akerib2018,Neves2017} and attribute the discrepancy to a different sample surface preparation or surface degradation.

In our detector the PTFE surfaces are not optically polished or treated, therefore we assume the surface is rougher and the diffuse component may play a more important role. Furthermore, the implementation of an angle-dependent mixture of diffuse and specular reflection components is not straightforward within the standard \textsc{Geant4} optical surface model. Considering both the experimental indications and the practical limitations of the implementation, we therefore adopted a purely diffuse reflection model as a reasonable approximation of the PTFE optical response.
\newline
\newline
\textbf{Optical properties of the materials}
\newline

The optical properties of liquid xenon relevant for photon transport were included in the simulation. The wavelength-dependent refractive index used in the simulation follows the measurements reported in Ref.~\cite{Baldini_2006}. The Rayleigh scattering length was set to 36 cm \cite{RayleighLXe}.

The absorption of scintillation photons in liquid xenon was neglected in the simulation. This approximation is justified by the very long attenuation length of liquid xenon at its own scintillation wavelength, which exceeds 10 m \cite{impurities}. Since this value is much larger than the characteristic dimensions of our detector, the effect is expected to be negligible.

The optical interface between liquid xenon and the quartz window placed in front of the SiPMs was also modeled. The quartz window transparency was set to 90$\%$, and its refractive index was taken as 1.6, according to the manufacturer specifications. Reflection and refraction at the LXe–quartz interface were handled by the optical boundary processes implemented in \textsc{Geant4}. The wavelength-dependent photodetection efficiency of the SiPMs was implemented according to the sensor datasheet.
\newline
\newline
\textbf{Evaluation of optical modeling uncertainties}
\newline

Several optical parameters entering the simulation carry non-negligible uncertainties. To evaluate the impact of these uncertainties on the simulated detector response, we performed dedicated simulations in which the relevant parameters were varied within reasonable ranges.

In particular, the refractive index of the quartz window protecting the SiPMs was varied between 1.55 and 1.65 around the nominal value of 1.6. The reflectivity of the PTFE reflectors was varied between 98$\%$, as reported in the literature, and 99.99$\%$ to account for the high reflectivity of PTFE in the VUV range. In addition, simulations were performed both including and excluding the liquid xenon layer located behind the quartz window in order to evaluate the effect of this region on photon transport.

For each variation of these parameters the full simulation chain was repeated and the resulting changes in the predicted detector response were evaluated. The spread of the results obtained from these simulations is used to estimate the systematic uncertainty associated with the optical transport modeling and is included in the Monte Carlo results in section \ref{meas}.
\newline

\acknowledgments
This work was supported by the European Research Council under grant ID 757829 and is part of the PRE2021-097277 grant, funded by MCIN/AEI/10.13039/501100011033 and the ESF+. We would like to warmly thank G. Llos\`a and R. Soleti for the barium and cobalt calibration sources used for the measurements.

\bibliography{biblio}

@misc{naI,
title="{{NaI(Tl)} properties}",
url="{https://www.crystals.saint-gobain.com/radiation-detection-scintillators/crystal-scintillators/naitl-scintillation-crystals#}",
note="{Online, accessed: 2023-03-30}"
}

@misc{csI,
title="{{CsI(Tl)} properties}",
url="{https://www.advatech-uk.co.uk/csi_tl.html}",
note="{Online, accessed: 2023-03-30}"
}

@misc{lysoCe,
title="{{LYSO(Ce)} properties}",
url="{https://www.advatech-uk.co.uk/lyso_ce.html}",
note="{Online, accessed: 2023-03-30}"
}

@misc{bgo,
title="{{BGO} properties}",
url="{https://www.advatech-uk.co.uk/bgo.html}",
note="{Online, accessed: 2023-03-30}"
}

@misc{hama_curve,
title="{Linearity curve for Hamamatsu SiPMs}",
url="{https://www.hamamatsu.com/eu/en/resources/interactive-tools/mppc-sipm-linearity.html}",
note="{Online, accessed: 2023-07-18}"
}

@misc{MaterialsFutureColliders,
  
  author = {Yeh, Minfang and Zhu, Ren-Yuan},
  
  title = "{Materials for Future Calorimeters}",
  
  publisher = {arXiv},
  
  year = {2022},
  
  copyright = {Creative Commons Attribution Non Commercial No Derivatives 4.0 International}
}

@article{review_fb,
author         = "Vandeberghe, S and Moskal, P and Karp, J",
journal = "EJNMMI Physics",
volume = "7",
year="2020",
number = "I",
pages = "35",
title = "{State of the art in total body PET}",
}

@article{Westerwoudt_2014,
	Author = {Victor Westerwoudt and Maurizio Conti and Lars Eriksson},
	Title = "{Advantages of Improved Time Resolution for {TOF PET} at Very Low Statistics}",
	Journal = {IEEE Trans. Nucl. Sci.},
	Pages = {126 - 133},
	Number = {1},
	Volume = {61},
	Year = {2014}
}

@article{Conti_2019,
	Author = {Maurizio Conti and Bernard Bendriem},
	Title = "{The new opportunities for high time resolution clinical {TOF PET}}",
	Journal = {Clinical and Translational Imaging},
	Pages = {139 - 147},
	Volume = {7},
	Year = {2019}
}

@article{Cherry_2018,
	Author = {Simon R. Cherry and others},
	Title = {Total-Body {PET}: Maximizing Sensitivity to Create New Opportunities for Clinical Research and Patient Care},
	Journal = {J. Nucl. Med.},
	Number = {1},
	Pages = {3-12},
	Volume = {59},
	Year = {2018}
}

@article{Badawi_2019,
	Author = {Ramsey D. Badawi and others},
	Title = {First human imaging studies with the {EXPLORER} total-body {PET} scanner},
	Journal = {J. Nucl. Med.},
	Pages = {299-303},
	Volume = {60},
	Year = {2019}
}

@article{Aprile:2010,
      author         = "Aprile, E. and Doke, T.",
      title          = "{Liquid xenon detectors for particle physics and astrophysics}",
      journal        = " Rev. Mod. Phys",
      volume         = "82",
      year           = "2010",
      pages          = "2053",
}

@article{FUJII2015293,
title = {High-accuracy measurement of the emission spectrum of liquid xenon in the vacuum ultraviolet region},
journal = {Nuclear Instruments and Methods in Physics Research Section A: Accelerators, Spectrometers, Detectors and Associated Equipment},
volume = {795},
pages = {293-297},
year = {2015},
author = {Fujii, K and others}
}

@article{Chepel:2012sj,
      author         = "Chepel, V. and Araujo, H.",
      title          = "{Liquid noble gas detectors for low energy particle
                        physics}",
      journal        = "JINST",
      volume         = "8",
      year           = "2013",
      pages          = "R04001",
}

@article{PhysRevB.17.2762,
  title = {Recombination luminescence in liquid argon and in liquid xenon},
  author = {Kubota, S. and Nakamoto, A. and Takahashi, T. and Hamada, T. and Shibamura, E. and Miyajima, M. and Masuda, K. and Doke, T.},
  journal = {Phys. Rev. B},
  volume = {17},
  issue = {6},
  pages = {2762--2765},
  year = {1978},
}

@article{Hogenbirk_2018,
   title={Field dependence of electronic recoil signals in a dual-phase liquid xenon time projection chamber},
   volume={13},
   number={10},
   journal={Journal of Instrumentation},
   author={Hogenbirk, E. and others},
   year={2018},
   pages={P10031–P10031}
}

@article{Gallucci_2009,
year = {2009},
volume = {160},
number = {1},
pages = {012011},
author = {Giovanni Gallucci},
title = {The {MEG} liquid xenon calorimeter},
journal = {Journal of Physics: Conference Series},
}

@article{PhysRevLett.123.161802,
  title = {Search for Neutrinoless Double-$\ensuremath{\beta}$ Decay with the Complete {EXO}-200 Dataset},
  author = {Anton, G. and others},
  collaboration = {EXO-200 Collaboration},
  journal = {Phys. Rev. Lett.},
  volume = {123},
  issue = {16},
  pages = {161802},
  year = {2019},
}

@article{xenon1T,
author = {Aprile, E. and others},
title = {Energy resolution and linearity of {XENON1T} in the MeV energy range},
journal = {Eur. Phys. J. C},
volume = {80},
pages = {785},
year = {2020},
}

@article{Adhikari_2021,
    title = "{nEXO: neutrinoless double beta decay search beyond 1028 year half-life sensitivity}",
	year = 2021,
	publisher = {{IOP} Publishing},
	volume = {49},
	number = {1},
	pages = {015104},
	author = {G Adhikari and others},
	journal = {Journal of Physics G: Nuclear and Particle Physics}
}

@article{Aalbers_2016,
	year = 2016,
	publisher = {{IOP} Publishing},
	volume = {2016},
	number = {11},
	pages = {017--017},
	author = {J. Aalbers and others},
	title = {{DARWIN}: towards the ultimate dark matter detector},
	journal = {Journal of Cosmology and Astroparticle Physics}
}

@article{Lavoie,
author = {L. Lavoie},
year = {1976},
journal = {Med. Phys.},
title = {Liquid xenon scintillators for imaging of positron emitters},
volume = {3 (5)},
pages = {283-293}
}

@article{chepelRes,
    author         = "Chepel, V. and others",
    title          = "{The liquid xenon detector for PET: recent results}",
    year           = "1999",
    journal        = "IEEE Transactions on Nuclear Science",
    volume         = "NS-46",
     pages          = "1038-1044",
}

@article{chepelEnergyRes,
    author         = "Crespo, P. and others",
    title          = "{Pulse processing for the PET liquid xenon Multiwire Ionisation Chamber}",
    year           = "1999",
    journal        = "IEEE Transactions on Nuclear Science",
    volume         = "47",
    issue = "6",
    pages          = "2119-2126",
}

@article{Doke1,
      author         = "Doke, T. and Kikuchi, J. and  Nishikido, F.",
      title          = "{Time-of-flight positron emission tomography using liquid xenon scintillation}",
      year           = "2006",
      journal        = "Nucl. Instrum. Methods A",
      volume         = "569",
      pages          = "863",
}

@article{miceli,
author ="Miceli, A",
title = "Liquid Xenon Detectors for Positron Emission Tomography",
journal = "J. Phys. Conf. Ser.",
volume = "312",
year = "2011",
pages = "062006",
}

@article{xemis_ER,
author ="Gallego Manzano, L and others",
title = "{XEMIS}: A liquid xenon detector for medical imaging",
journal = "{Nucl. Instrum. Meth.}",
volume = "{A787}",
pages = "89--93",
year = "2015"
}

@article{ChepelEres,
      author         = "Chepel, V",
      title = "Liquid Xenon Detectors for Medical Imaging",
      journal        = "Revista do Detua ",
      volume         = "4",
      number         = "7",
      year           = "2007"
}

@article{Gomez-Cadenas:2016mkq,
    author = "Gomez-Cadenas, J. J and others",
    title = "{Investigation of the Coincidence Resolving Time performance of a PET scanner based on liquid xenon: A Monte Carlo study}",
    journal = "JINST",
    volume = "11",
    number = "09",
    pages = "P09011",
    year = "2016"
}

@article{Gomez-Cadenas:2017bfq,
    author = "Gomez-Cadenas, Juan Jose and Benlloch-Rodríguez, José María and Ferrario, Paola",
    title = "{Monte Carlo study of the Coincidence Resolving Time of a liquid xenon PET scanner, using Cherenkov radiation}",
    journal = "JINST",
    volume = "12",
    number = "08",
    pages = "P08023",
    year = "2017"
}

@inproceedings{IEEE2018talk,
author="Herrero-Bosch, V and others",
title = "{PETALO read-out: A novel approach for data acquisition systems in PET applications}",
booktitle = "{2018 IEEE Nuclear Science Symposium and Medical Imaging Conference (NSS/MIC)}",
year = "2018"
}

@inproceedings{Renner:2020ayj,
    author = "Renner, J and others",
    title = "{Processing of Compton events in the PETALO readout system}",
    booktitle = "{2019 IEEE Nuclear Science Symposium (NSS) and Medical Imaging Conference (MIC)}",
    pages = "1--7",
    year = "2019"
}

@inproceedings{Ferrario:2019pvw,
    author = "Ferrario, Paola and others",
    title = "{PETALO: Time-of-Flight PET with liquid xenon}",
    booktitle = "{2018 IEEE Nuclear Science Symposium and Medical Imaging Conference (NSS/MIC)}",
    month = "11",
    year = "2018"
}

@article{Renner_2022,
	year = 2022,
	month = {may},  
	publisher = {{IOP} Publishing},  
	volume = {17},  
	pages = {P05044},  
	author = {J. Renner and others}, 
	title = {Monte Carlo characterization of {PETALO},
   a full-body liquid xenon-based {PET} detector}, 
	journal = {Journal of Instrumentation}
}

@article{Nerea_2025,
    title = {Evaluation of coincidence time resolution in a liquid xenon detector with silicon photomultipliers},
  author = {Salor-Igui\~niz, N and others},
  journal = {Phys. Rev. Res.},
  volume = {7},
  issue = {3},
  pages = {033089},
  year = {2025},
}

@article{LZ,
  title = {Energy resolution of the {LZ} detector for high-energy electronic recoils},
  author = {Pereira, G. and Silva, C. and Solovov, V. N.},
  journal = {JINST},
  volume = {8},
  year = {2023},
  issue = {04},
  pages = {C04007},
}

@article{non-prop,
author = { Tadayodshi   Doke  and  Ryu   Sawada  and  Hiroko   Tawara },
title = {NON-PROPORTIONALITY OF THE SCINTILLATION YIELD IN LIQUID XENON AND ITS EFFECT ON THE ENERGY RESOLUTION FOR GAMMA-RAYS},
journal = {Technique and Application of Xenon Detectors},
year = {2003},
pages = {17-27},
}

@article{nest,
    title={Noble Element Simulation Technique v2.0 (Version v2.0.0)},
    journal={Zenodo},
   author={Szydagis, M and others},
   year={2018},

}

@article{YAMASHITA2004692,
title = {Scintillation response of liquid Xe surrounded by PTFE reflector for gamma rays},
journal = {Nuclear Instruments and Methods in Physics Research Section A: Accelerators, Spectrometers, Detectors and Associated Equipment},
volume = {535},
number = {3},
pages = {692-698},
year = {2004},
author = {M. Yamashita and others}
}

@article{PhysRevC.84.045805,
  title = {New measurement of the scintillation efficiency of low-energy nuclear recoils in liquid xenon},
  author = {Plante, G. and Aprile, E. and Budnik, R. and Choi, B. and Giboni, K.-L. and Goetzke, L. W. and Lang, R. F. and Lim, K. E. and Melgarejo Fernandez, A. J.},
  journal = {Phys. Rev. C},
  volume = {84},
  issue = {4},
  pages = {045805},
  year = {2011},
  publisher = {American Physical Society},
}

@article{Ni:2006zp,
      author         = "Ni, K. and others",
      title          = "{Gamma Ray Spectroscopy with Scintillation Light in
                        Liquid Xenon}",
      journal        = "JINST",
      volume         = "1",
      year           = "2006",
      pages          = "P09004",
}

@article{impurities,
title = {Absorption of scintillation light in a 100 l liquid xenon gamma-ray detector and expected detector performance},
journal = {Nuclear Instruments and Methods in Physics Research Section A},
volume = {545},
pages = {753-764},
year = {2005},
author = {Baldini, A and others},
}

@article{AKERIB201334,
title = {Technical results from the surface run of the LUX dark matter experiment},
journal = {Astroparticle Physics},
volume = {45},
pages = {34-43},
year = {2013},
author = "Akerib, D.S. and others"
}

@article{petsys,
title = {Tofpet2: a high- performance asic for time and amplitude measurements of sipm signals in time-of-flight applications},
journal = {JINST},
volume = {11},
number = {3},
pages = {C03042},
year = {2016},
author = {Di Francesco, A and others}
}

@article{Geant,
	Author = {S. Agostinelli and others},
	Title = {{GEANT}4 - a simulation toolkit},
	Journal = {Nucl. Instrum. Meth. A},
	Pages = {250},
	Volume = {506},
	Year = {2003}
}

@article{Baldini_2006,
	year = 2006,
	publisher = {Institute of Electrical and Electronics Engineers ({IEEE})},
	volume = {13},
	number = {3},
	pages = {547--555},
	author = {A. Baldini and others},
	title = {Liquid xenon scintillation calorimetry and Xe optical properties},
	journal = {{IEEE} Transactions on Dielectrics and Electrical Insulation}
}

@article{RayleighLXe,
      author         = "Solovov, V.N. and others",
      title = "Measurement of the refractive index and attenuation length of liquid xenon for its scintillation light",
      journal        = "Nucl. Instrum. Meth. A ",
      volume         = "516",
      pages = "462",
      year           = "2004",
}

@article{Silva2010,
      author         = "C. Silva and others",
      title = "Reflectance of polytetrafluoroethylene for xenon scintillation light.",
      journal        = "J. Appl. Phys",
      volume         = "107",
      number         = "6",
      pages = "064902",
      year           = "2010",
}

@article{EpjcPTFE,
      author   = "Kravitz, S. and others",
      title    = "{Measurements of angle-resolved reflectivity of PTFE in liquid xenon with IBEX}",
      journal  = "The European Physical Journal C",
      volume   = "80",
      number   = "3",
      pages    = "262",
      year     = "2020",
}

@article{Akerib2018,
  title = "{Calibration, event reconstruction, data analysis, and limit calculation for the LUX dark matter experiment}",
  author = {Akerib, D. S.  and others},
  collaboration = {LUX Collaboration},
  journal = {Phys. Rev. D},
  volume = {97},
  issue = {10},
  pages = {102008},
  year = {2018},
  publisher = {American Physical Society},
}

@article{Neves2017,
  title = "{Measurement of the absolute reflectance of polytetrafluoroethylene (PTFE) immersed in liquid xenon}",
  author = {F. Neves  and others},
  journal = {JINST},
  volume = {12},
  pages = {P01017},
  year = {2017},
}

\end{document}